\documentclass{aa}
\usepackage{natbib}
\usepackage{color}
\usepackage{graphicx}
\usepackage[varg]{txfonts}
\usepackage{hyperref}
\hypersetup{citecolor=black, urlcolor=blue, linkcolor=black, colorlinks=true}
\newcommand{\sect}[1]{Sect.\,\ref{#1}}

\newcommand{\fig}[1]{Fig.\,\ref{#1}}
\newcommand{\figs}[1]{Figs.\,\ref{#1}}

\definecolor{orange}{rgb}{1,0.4,0.}
\graphicspath{{./}{figs-ps/}{figs/}}
\begin{document}

%=============================================================================
% TITLE
%=============================================================================
%
\title{Transient small-scale brightenings in the quiet solar corona: A model for campfires observed with Solar Orbiter} 
\titlerunning{campfire-model}

%\authorrunning{}

\author{Yajie Chen\inst{1,2}
          \and
          Damien Przybylski\inst{2}
          \and
          Hardi Peter\inst{2}
          \and
          Hui Tian\inst{1,3}
          \and
          F. Auch\`ere\inst{4}
          \and
          D. Berghmans\inst{5}
          }

\institute{School of Earth and Space Sciences, Peking University, 100871 Beijing, China\\
              \email{chenyajie@pku.edu.cn; huitian@pku.edu.cn}
         \and
             Max-Planck Institute for Solar System Research, 37077 G\"{o}ttingen, Germany
         \and
             Key Laboratory of Solar Activity, National Astronomical Observatories, Chinese Academy of Sciences, Beijing 100012, China
        \and 
            Universit\'e Paris-Saclay, CNRS, Institut d’Astrophysique Spatiale, 91405 Orsay, France
        \and 
            Solar-Terrestrial Centre of Excellence – SIDC, Royal Observatory of Belgium, 1180 Brussels, Belgium
          }

\date{Version: \today}
%\date{Received xx Jan 2021 / Accepted xx yyy 2021}

  \abstract
  {Recent observations by the Extreme Ultraviolet Imager (EUI) on board Solar Orbiter have characterized prevalent small-scale transient brightenings in the corona above the quiet Sun termed campfires.}
  {In this study we search for comparable brightenings in a numerical model and then investigate their relation to the magnetic field and the processes that drive these events.}
  {We used the MURaM code to solve the 3D radiation magnetohydrodynamic equations in a box that stretches from the upper convection zone to the corona. The model self-consistently produces a supergranular network of the magnetic field and a hot corona above this quiet Sun. For the comparison with the model, we synthesized the coronal emission as seen by EUI in its 174\,\AA\ channel, isolated the seven strongest transient brightenings, and investigated the changes of the magnetic field in and around these in detail.}
  {The transients we isolated have a lifetime of about 2 minutes and are elongated loop-like features with lengths around 1\,Mm to 4\,Mm. They tend to occur at heights of about 2\,Mm to 5\,Mm above the photosphere, a bit offset from magnetic concentrations that mark the bright chromospheric network, and they reach temperatures of above 1\,MK. As a result, they very much resemble the larger campfires found in observations.
  In our model most events are energized by component reconnection between bundles of field lines that interact at coronal heights.
  % and do not show signatures of flux emergence or cancellation in the photosphere.
  In one case, we find that untwisting a highly twisted flux rope initiates the heating.}
  {Based on our study, we propose that the majority of campfire events found by EUI are driven by component reconnection and our model suggests that this process significantly contributes to the heating of the corona above the quiet Sun.}

%\abstract{Recent observations by the Extreme Ultraviolet Imager (EUI) on board Solar Orbiter have revealed prevalent small-scale transient brightenings in the quiet solar corona termed "campfires". To understand the generation mechanism of these coronal brightenings, we constructed a self-consistent and time-dependent quiet-Sun model extending from the upper convection zone to the lower corona using a realistic three-dimensional radiation magnetohydrodynamic simulation. From the model we have synthesized the coronal emission in the EUI 174 {\AA} passband. We identified several transient coronal brightenings similar to those in EUI observations. The size and lifetime of these coronal brightenings are mostly 0.5--4 Mm and $\sim$2 min, respectively. These brightenings are generally located at a height of 2--4 Mm above the photosphere, and the local plasma is often heated above 1 MK. By examining the magnetic field structures before and after the occurrence of brightenings, we concluded that these coronal brightenings are generated by component magnetic reconnection between interacting bundles of magnetic field lines or neighboring field lines within highly twisted flux ropes. Occurring in the coronal part of the atmosphere, these events generally reveal no obvious signature of flux emergence or cancellation in photospheric magnetograms. These transient coronal brightenings may play an important role in heating of the local coronal plasma.}

%
%----------------------------------------------------------------------------

\maketitle

%==============================================================================
\section{Introduction\label{S:intro}}
%==============================================================================
On average, in the solar atmosphere, the temperature increases with height above the chromosphere and reaches $\sim$1 MK in the corona. Clearly, the heating of the upper atmosphere is related to the magnetic field, but how the energy is generated, transported, and dissipated is still under debate \citep[e.g.,][]{Aschwanden2019}.
Observations in the extreme ultraviolet (EUV) have been used abundantly to study the statistics of heating events as well as individual events to infer properties of the underlying heating process(es).

In terms of imaging, the  Extreme-ultraviolet Imaging Telescope \citep[EIT,][]{Delaboudini1995}, the Transition Region and Coronal Explorer \citep[TRACE,][]{Handy1999}, the Extreme UltraViolet Imager \cite[EUVI;][]{Howard2008}, and the Atmospheric Imaging Assembly \citep[AIA,][]{Lemen2012} have been used extensively.
The latest of these, AIA, achieved a spatial resolution of about 1.4\arcsec\ or 1000\,km on the Sun (at disk center).
{In addition, recent rocket flights of the High-resolution Coronal imager \citep[Hi-C;][]{HiC,HiC-2.1} have achieved a spatial resolution of about 0.4\arcsec to 0.5\arcsec, corresponding to about 300 to 350\,km on the Sun.}
These instruments have been used to investigate the distribution of events in terms of energy deposition, or rather radiative output, and they generally found that this follows a power law with most events happening at small spatial scales \cite[e.g.,][]{Berghmans1998,Harra2000,Aschwanden2002, Tiwari2019}.
The larger brightenings, for example coronal bright points, that have been spatially resolved often show a temporal evolution, as is expected for reconnection events \cite[e.g.,][]{Madjarska2019}.
The role of reconnection for transient brightenings is underlined by spectroscopic investigations.
These show observational signatures of reconnection in the form of bi-directional flows or extreme line broadening in small-scale brightenings, may it be explosive events \citep[e.g.,][]{Dere1989,EEs,Chen2019b} or UV bursts \cite[e.g.,][]{Peter2014,Tian2018,Young2018}.

{Such reconnection brightenings observed in (E)UV} have been studied in numerous 2D and 3D numerical models based on magnetohydrodynamics (MHD).
The 2D models have mostly been tailored to reproduce the observational signatures of reconnection, in particular for explosive events, either with an idealized setup of a current sheet \citep[e.g.,][]{Innes2015, Ni2016} or for cases where the magnetic structure is driven from the surface \citep[e.g.,][]{Peter2019,Ni2020}.
More comprehensive 3D models could even give a self-consistent description of UV bursts \cite[][]{Hansteen2017,Hansteen2019}.
In essentially all of these numerical models {about explosive events and UV bursts}, inducing both 2D and 3D models, the reconnection occurs between almost antiparallel magnetic field structures.

Recently, the Extreme Ultraviolet Imager \citep[EUI;][]{EUI} on board Solar Orbiter \citep[][]{SolO} performed observations of the quiescent corona in the 174 {\AA} passband near the disk center. These data were taken during the first perihelion pass of Solar Orbiter at a distance of $\sim$0.6 AU from the Sun while still being in the commissioning phase. 
The High-Resolution Imager (HRI) of EUI provided EUV images at a very high spatial resolution of about 400\,km on the Sun. The whole observation sequence with a cadence of 5\,s lasted for less than five minutes, but this was sufficient to study more than a thousand compact loop-like brightenings, now termed campfires \cite[][]{campfires}. The observational results concerning these campfires can be summarized as follows \cite[based on][]{campfires}:

\begin{enumerate}
    \item The distribution of lifetimes is power-law-like with most of the events observed by EUI/HRI lasting for less than 2 minutes.
    \item The bulk of the detected events have linear sizes of below 4\,Mm and the sizes follow a power-law-like distribution.
    \item Most events are elongated loop-like features with an aspect ratio of length to width ranging from 1 to 5.
    \item They occur at the edges of bright patches of the chromospheric network, that is,\ close to but not directly above the brightest parts of the network. 
    \item Triangulation with AIA data revealed a height of these events in the range of 1\,Mm to 5\,Mm above the photosphere. 
    \item They reach temperatures between 1\,MK and 1.6\,MK.
    \item The total radiative losses in the 174\,\AA\ channel also follow a power-law-like distribution. 
\end{enumerate}

Aspects of these findings have been reported before. For example, \cite{Chitta2021} have recently reported on statistics of small-scale brightenings observed with AIA, which are probably similar in nature to the campfires, but they appear on larger scales. This is clear because the spatial resolution of AIA on the Sun is about a factor of 2.5 worse than EUI at 0.6\,AU. A spatial resolution comparable to, or slightly better than, EUI has been achieved by Hi-C. With 0.4\arcsec\ to 0.5\arcsec, Hi-C could resolve structures down to 300\,km to 350\,km; of course, if it were at future perihelia closer than 0.3\,AU to the Sun, EUI would surpass Hi-C. Similar to the campfires, Hi-C could identify tiny loops with lengths as small as 1\,Mm and with lifetimes, aspect ratios, and a location with respect to the network similar to campfires \cite[][]{Peter2013,Barczynski2017}. However, only a handful of these miniature loops have been reported. 
Observations from IRIS \citep{IRIS} and Hi-C have also revealed numerous subarcsec brightenings \citep[e.g., ][]{Tian2014,Alpert2016}. However, these elongated brightenings are mostly seen in transition region lines and no clear coronal response was detected.
Combining the results 1 to 7 as listed above, EUI described a feature which, in its completeness, escaped detection through previous instruments.

Our study aims to see if transient coronal brightenings similar to the campfires are also found in a 3D model, and if so, to understand their nature. For this, we constructed a self-consistent and time-dependent quiet-Sun model extending from the upper convection zone to the corona using a realistic 3D radiation MHD simulation. We synthesized EUI 174 {\AA} images from the model, that is, as EUI would see our model. Finding campfire-like events, we investigated their thermal properties and associated magnetic field topologies. Our study suggests that these transient brightenings are mostly associated with plasma heating due to component reconnection.

%==============================================================================
\section{MHD model and coronal emission\label{S:model}}
%==============================================================================

%------------------------------------------------------------------------------
\subsection{Radiation MHD simulation using MURaM\label{S:muram}}
%------------------------------------------------------------------------------

We performed a 3D radiation MHD simulation using the coronal extension of the MURaM code \citep{muram2005, Rempel2017}. The computational domain covers a region of 50$\times$50 Mm$^{2}$ in the horizontal direction and extends from $\sim$20 Mm below to $\sim$17.5 Mm above the photosphere in the vertical direction. The grid size of the simulation is $\sim$48.8 km and 25 km in the horizontal and vertical directions, respectively.

We used periodic boundary conditions in the horizontal direction. The upper boundary condition is open to outflows, closed to inflows, and the magnetic field is specified by a potential boundary condition. The lower boundary condition is open to inflows and outflows, with symmetric magnetic field components \cite[Open-boundary Symmetric-field (OSb) in][]{Rempel2014}. {This boundary condition allows inflows to carry magnetic fields near the equipartition field strength into the domain in order to model recirculation deeper in the convection zone. Although a small-scale dynamo can operate locally in the photosphere \citep{schussler_2008_dynamo}, a boundary condition that models a deeper convection zone is required to reproduce the observationally inferred quiet-Sun field strengths \citep{Rempel2014}.}

The simulation parameters are similar to those described in \citet{Rempel2017} with a few modifications. For the higher resolution simulations presented in this work, the diffusion h-parameters were decreased slightly to 1.8 in the photosphere and chromosphere (density $\rho{>}10^{-10}$\,g\,cm$^{-3}$). In the corona ($\rho{<}10^{-10}$\,g\,cm$^{-3}$),  a value of $h{=}0.8$ was used for the diffusion of the density, energy, velocity, and $h{=}3.2$ on the magnetic field. We used gray local-thermal-equilibrium (LTE) radiative transfer in the photosphere and low chromosphere. We used a pre-tabulated equation of state generated using FreeEoS \citep{Irwin_2012} and opacities from MPS-ATLAS \citep{MPS_ATLAS_WITZKE}.  From the mid-chromosphere to the corona, we used the tabulated H, Ca, and Mg line losses as well as the optically thin losses described by \citet{Carlsson2012}. In order to prevent double cooling in the chromosphere, we turned off the 3D radiation scheme at an optical depth of $\tau_{\rm{cutoff}}=10^{-5}$ with a function of the form $\tau_{5000}^2 / \left(\tau_{5000}^2 + \tau_{\rm{cutoff}}^2\right)$.

The simulation was initialized with the energy and density profiles from a G2V stellar model, which is constant in the horizontal directions.  The simulation initially extends $\sim$1 Mm above the solar surface. A small random velocity field was used to start convection. This setup was run for 53 hours of solar time until convection developed throughout the box. A zero net-flux magnetic field seed field was then added by setting a random vertical field for each horizontal pixel, with a root mean square (RMS) value of $10^{-3}$\,G, and it is constant in the vertical direction. This setup was run for another 58 hours until the magnetic field strength was saturated. The simulation was then extended 5 Mm further above the surface and run for an hour until a stable transition region formed. Finally, the domain was extended another 11.5 Mm and run for 30 minutes until the corona was self-consistently heated. We collected snapshots for 16 minutes at a cadence of 20\,s.

%------------------------------------------------------------------------------
\subsection{Coronal emission synthesized from the model\label{S:emission}}
%------------------------------------------------------------------------------

In order to compare our model results with actual coronal observations, we synthesized coronal emission in the EUI 174\,{\AA} and AIA 171\,{\AA} channels by using response (or contribution) functions $R(T,n)$ that mainly depend on temperature $T$ and to a lesser degree on electron density $n$.
In the optically thin corona, where the lines (and continua) are excited by electron collisions, the radiative loss (per unit volume) is then given by $L=n^2\,R(T,n)$.

\begin{figure} 
%\centering {\includegraphics[width=88mm]{figS4.eps}} 
\centering {\includegraphics[width=88mm]{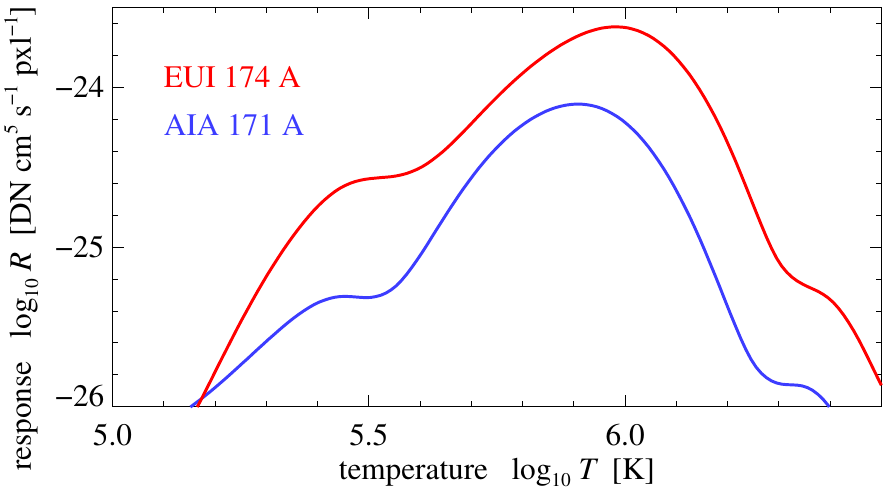}}
\caption{Temperature response, or contribution, function $R(T,n)$.
The red line shows the response for the EUI channel at 174\,{\AA}, and the blue line corresponds to the one for AIA at 171\,{\AA}.
Both responses are shown here for a density of $n=10^{9}$\,cm$^{-3}$, which is typical for coronal loops.
The radiative losses (per volume) in the respective band are given by $n^2R$.
See \sect{S:emission}.
} \label{F:response}
\end{figure}

To determine $R(T,n)$, we first calculated the spectra in the respective wavelength regions using Chianti version\,10 \cite[][]{Chianti,Chianti-v10} with the standard Chianti ionization equilibrium and the coronal abundances of \cite{Feldman1992}.
We did this for a grid of temperatures and densities; additionally, for each combination of temperatures and densities, we multiplied the resulting spectrum with the effective area (as a function of wavelength) of the respective channel. Integrating over the wavelength with the proper normalization then provides the response function $R(T,n)$.
We show the response functions for the EUI 174\,{\AA} and AIA 171\,{\AA} channels in \fig{F:response}.
In the case of the AIA response, we also considered the degradation of its sensitivity, which is why the AIA response is weaker than the EUI response.
The peak in EUI is at slightly higher temperatures because the EUI 174\,\AA\ band is dominated by a mix of  \ion{Fe}{ix} and \ion{Fe}{x}, while the AIA 171\,\AA\ band is dominated by \ion{Fe}{ix}. Also, the contribution from the transition region around a few $10^5$\,K is slightly more pronounced in EUI.
Otherwise the two response curves are very similar.
The EUI/HRI contribution function used in this paper is based on a preliminary analysis of the calibration data. The absolute value and wings will likely change somewhat with the final preflight calibration. The central wavelength and overall shape of the peak are, however, not likely to be affected.

%>>>>>>>>>>>>>>>>>>>>>>>>>>>>>>>>>>>>>>>>>>>>>>>>>>>>>>>>>>>>>>>>>>>>>>>>>>>>>>
%<<<<<<<<<<<<<<<<<<<<<<<<<<<<<<<<<<<<<<<<<<<<<<<<<<<<<<<<<<<<<<<<<<<<<<<<<<<<<<

We calculated the radiative loss $L$ per unit volume at each grid cell by calculating $R(T,n)$ at that cell and multiplying it with $n^2$. To calculate the image as it would be seen by either EUI or AIA, we then simply integrated along the line of sight. In this study we focus on a vertical line of sight, that is, the images we show correspond to actual observations at the disk center. This is because we want to compare our model to the observations presented by \cite{campfires} that were taken at the disk center.

For better comparison to actual data, we also degrade the coronal images we synthesized from the model to roughly match the spatial resolution of the EUI High Resolution Imager (HRI) at 174\,{\AA} and the AIA channel at 171\,\AA.
In order to do this, we simply convolved the maps from the model with a symmetric 2D Gaussian.
For the kernels for EUI and AIA, we chose a full width at half maximum of 450\,km and 900\,km, respectively, which roughly corresponds to the resolution of the two instruments.
At the time of the campfire observations by \cite{campfires}, Solar Orbiter was about 0.6\,AU from the Sun. Then two EUI/HRT pixels correspond to about 400\,km, so our resolution estimate for EUI here is conservative.

%==============================================================================
\section{Results\label{S:results}}
%==============================================================================

\begin{figure*} 
\sidecaption
\centering {\includegraphics[width=12cm]{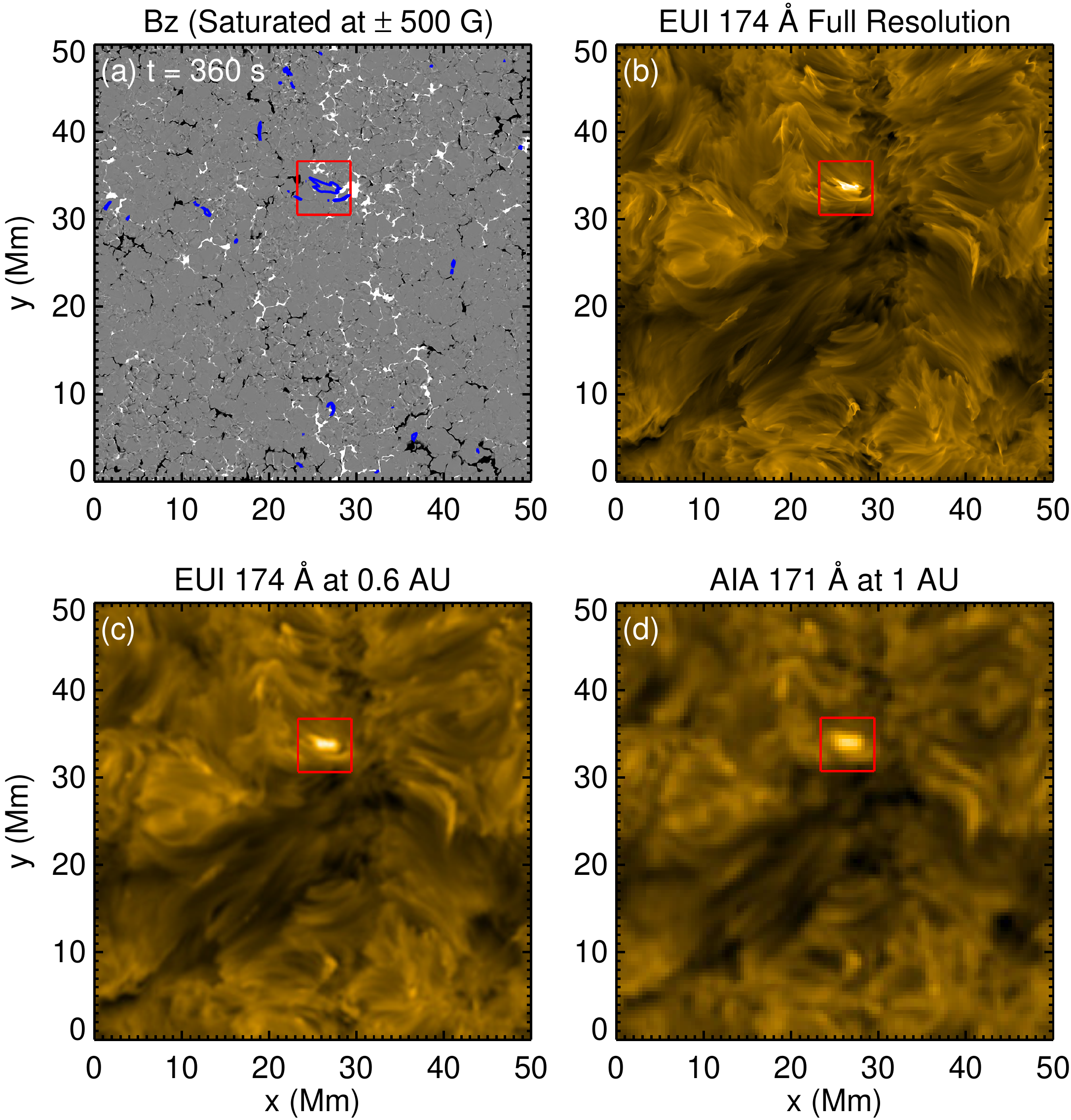}} 
%\centering {\includegraphics[width=16cm]{fig1.eps}} 
\caption{Photospheric magnetic field and coronal emission in a 3D MHD model.
Panel (a) shows the vertical component of the photospheric magnetic field saturated at ${\pm}$500 G.
This would correspond to a magnetogram at the disk center.
Panel (b) displays the coronal emission in the 174 {\AA} band integrated along the vertical direction at the full resolution of the numerical model (50\,km grid spacing in the horizontal directions).
Contours of regions with an EUV intensity of 5$\sigma$ above the average intensity are overplotted in blue on the magnetogram in panel (a) in order to relate bright patches to the magnetic structure.
Panel (c) shows the same emission in the EUI band as in panel (b), but now degraded to a spatial resolution similar to EUI at 0.6\,AU (ca. 200\,km per pixel plate scale).
For comparison, we show in panel (d) how AIA would see the model at 171\,\AA, that is, we reduced the resolution to match AIA (plate scale of ca. 450\,km per pixel).
The EUV images are shown on a logarithmic scale over a dynamic range of 1000. 
These panels show the full horizontal extent of the computational domain at one single snapshot.
The red box highlights one coronal brightening that is discussed in more detail in \fig{f2}.
An animation is available online.
See \sect{S:part1}.
%
%{\newline\color{red} IDEA FOR MOVIE: in the movie you could highlight the seven campfires for those snapshots we see them, i.e. during the red part of the light curves in \fig{fs2}. Each of the campfires would then be highlighted by a red box (as now in this figure) with the FOV of the detailed \figs{f3}, \ref{f4}, and \ref{fs2}, but  only for the snapshots they are seen. I do not think this needs to be done now, but maybe as an enhancement for the revision (so this can wait until after submission; it will take some manual work).}
%
} \label{f1}
\end{figure*}

In order to study transient coronal brightenings and their relation to the magnetic field, we first inspected the coronal images of the 16-minute time series of our numerical model for such events.
In \fig{f1} we display a single snapshot that shows the vertical magnetic field at the photosphere, that is,\ a magnetogram (panel a) and the corresponding coronal image in the 174\,\AA\ channel (panel b). Here, one brightening is highlighted by the red box that sits next to a small magnetic field concentration.

From the time series of the synthesized coronal images, we singled out seven individual events for a detailed case study (see \sect{S:part1}).
Of course, the criterion we chose there is somewhat arbitrary, but we wanted to select a few individual events that we can still study carefully on a case-by-case basis.
A future study should look at this, or a similar, model in a statistical way, for example,\ using the same procedures as in the campfire study by \cite{campfires}. Consequently, we were not be able to recover the power-law-like distributions as found in \cite{campfires}. Instead, we unravel the nature of a selection of the events by investigating changes in the magnetic field during each of the seven events.

In the following, we first compare the properties of the brightenings in our 3D model to the observed campfires (\sect{S:part1}).
We then investigate the magnetic nature of these brightenings in \sect{S:part2}

%>>>>>>>>>>>>>>>>>>>>>>>>>>>>>>>>>>>>>>>>>>>>>>>>>>>>>>>>>>>>>>>>>>>>>>>>>>>>>>
%<<<<<<<<<<<<<<<<<<<<<<<<<<<<<<<<<<<<<<<<<<<<<<<<<<<<<<<<<<<<<<<<<<<<<<<<<<<<<<

%------------------------------------------------------------------------------
\subsection{Magnetic fields in the photosphere\label{S:part0}}
%------------------------------------------------------------------------------

The photospheric magnetograms in our model show small-scale magnetic concentrations with a field strength on the order of one thousand Gauss and weak salt-and-pepper magnetic elements with a field strength of several hundred Gauss (see \fig{f1}a). This is consistent with observations of the quiet-Sun magnetic fields \citep[e.g.,][]{Rubio2019}. The magnetic fields in our model are maintained through a small-scale dynamo process \citep{Rempel2014}. The magnetic field concentrations form a long-living network pattern.
%, similar in appearance and in cell size to actual observations. 
The generation of network and internetwork fields are associated with convection on the supergranule and granule scales. With a horizontal extent of 50$\times$50\,Mm$^2$, our computational domain is large enough so that several supergranular cells fit into the box.

%------------------------------------------------------------------------------
\subsection{Coronal transient brightenings in the model\label{S:part1}}
%------------------------------------------------------------------------------

%The spatially averaged temperature in the coronal part of our quiet Sun model ranges from 1\,MK to 2\,MK,
{The spatially averaged temperature in the coronal part of our quiet-Sun model is $\sim$1.2 MK and remains stable in time. However, the temperature shows a large spatial variation due to the localization of transient heating events. The local temperature also reveals a large temporal variation because of the intermittency of the energy releases.}
Consequently, we find significant fluctuations in the coronal emission as synthesized in the 174\,{\AA} band in both space and time.

In order to identify transient brightenings as campfires, we applied the following procedure to the synthetic EUI images reduced to the approximate EUI/HRI resolution.
The intensity in a given spatial pixel (in the x-y intensity map) has to be $5\sigma$ above the mean value over the whole time series.
Furthermore, this has to be the case in at least nine contiguous pixels and for at least two snapshots (that are 20\,s apart).
If these conditions are met, we call this a campfire, which is the case for seven events in our model data.

\begin{figure*} 
\sidecaption
{\includegraphics[width=12cm]{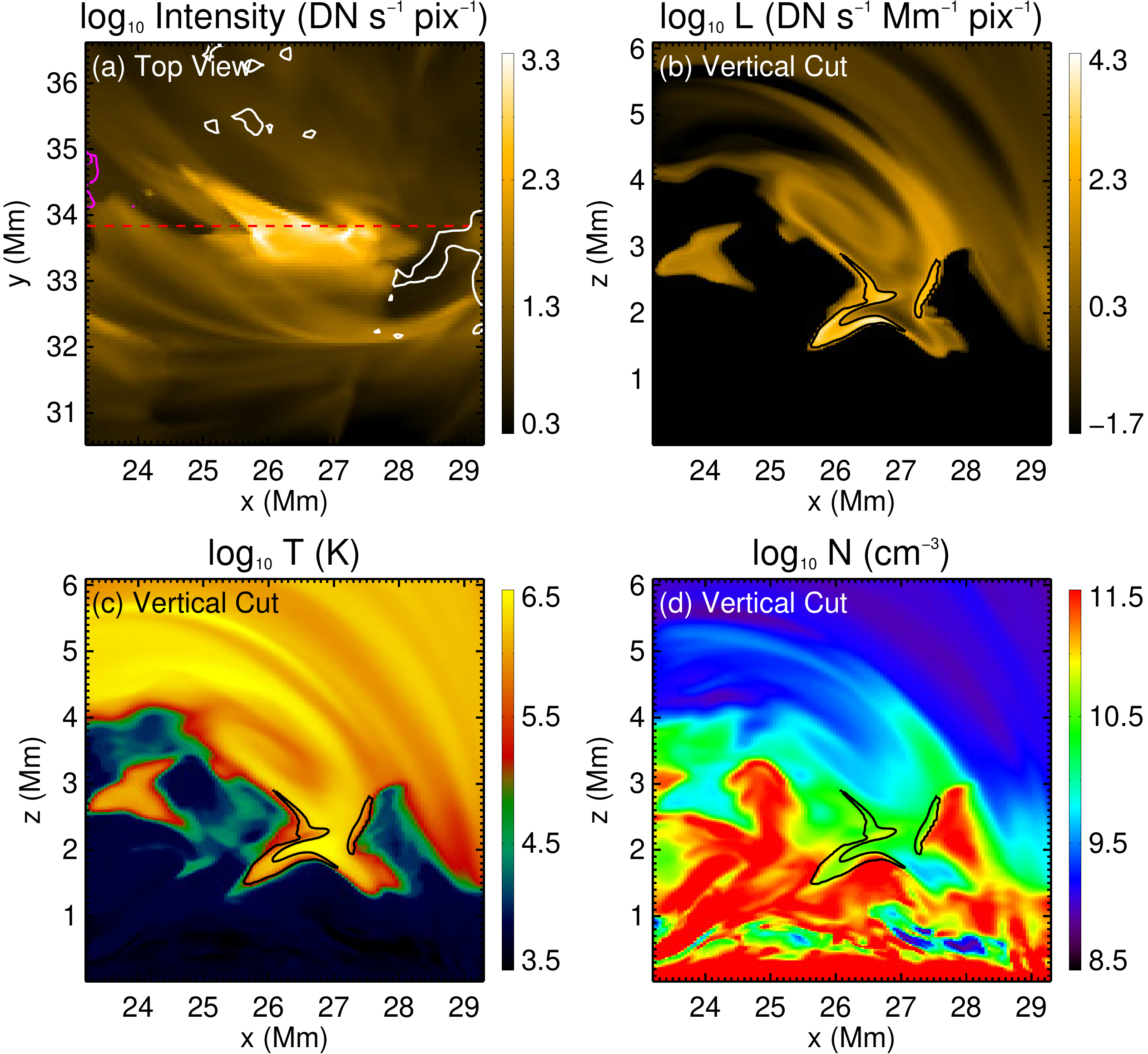}}
\caption{Thermal structure in a modeled campfire.
Panel (a) shows the synthesized emission in the EUI 174 {\AA} channel as seen from above. The field of view corresponds to the red box in Figure~\ref{f1}. The purple and white contours correspond to vertical magnetic fields at $\pm$1000\,G in the photosphere. Panels (b) to (d) display vertical cuts through the campfire at the y position indicated by the dashed line in panel (a). The cut of the radiative losses per volume in the 174\,\AA\ band is plotted in panel (b). The black contours outline the regions of enhanced radiation. These contours are also overplotted on the temperature (panel c) and number density (panel d) in order to read off the temperature and density in the campfire.
The event shown here is the same as the one shown in \fig{f3}c. The temporal evolution of the emission from this campfire is displayed in panel (4c) of \fig{fs2}.
See \sect{S:part1}.
}\label{f2}
\end{figure*}

We first concentrate on one of these seven events that is highlighted in \fig{f1}.
In \fig{f2}a we show a zoom into this brightening.
At the original model resolution, it has a flame-like morphology with a length of about 4\,Mm and a width of roughly 1\,Mm, and it lasts for about 2 minutes.
This is largely consistent with the size, aspect ratio, and lifetime of observed campfires, though this example would be on the large and long-living side (items 1, 2, and 3 in the introduction).

Comparing the location of the sample brightening highlighted in \fig{f1} to the location of the underlying magnetic field shows that it is offset from the magnetic concentration.
The zoom in \fig{f2}a underlines that the elongated brightening connects the underlying opposite magnetic polarities (shown as contours), but it does not reach them.
Therefore, we can also expect this campfire to be offset to the bright chromospheric network because the bright network patches are found directly above the strong magnetic field concentrations \cite[e.g.,][]{Barczynski2018}.
In \fig{f1}a, we mark the coronal 174\,\AA\ emission in the magnetogram by contour lines to show how the more numerous smaller brightenings are related to the magnetic field.
Just as for the bigger event, these avoid the centers of the network elements, that is, the magnetic concentrations.
This is because the field lines are rooted in the magnetic concentrations, but the coronal brightenings happen away from the photospheric footpoints and hence are offset when looking from straight above.
This fits well with the location of the observed campfires (item 4 in the introduction).

The brightening of the sample campfire from \fig{f1} occurs at a height of ca. 2\,Mm above the photosphere. This can be seen from the vertical cut in \fig{f2}b.
Other events we looked at in more detail show the bright emission at slightly higher altitudes, for example, the example in \fig{f4} occurs at about 4\,Mm.
Thus the campfires we find in our model occur in the same height range as found in the observations (item 5 in the introduction).

%>>>>>>>>>>>>>>>>>>>>>>>>>>>>>>>>>>>>>>>>>>>>>>>>>>>>>>>>>>>>>>>>>>>>>>>>>>>>>>
%<<<<<<<<<<<<<<<<<<<<<<<<<<<<<<<<<<<<<<<<<<<<<<<<<<<<<<<<<<<<<<<<<<<<<<<<<<<<<<

To examine the thermal properties of the sample campfire, we studied vertical cuts through this transient.
In \fig{f2} we show the synthesized EUI 174 {\AA} emission, temperature, and density through the middle of the brightening.
The temperature and density within the brightening are about 1\,MK and $3{\times}10^{10}$\,cm$^{-3}$.
Some plasma embedded in the bright structure was heated to significantly higher temperatures of 3 MK.
Because of 
%the lower density there and
the sharp drop in the response function above 1\,MK, this does not show up in 174 {\AA} emission.
Still, it might show up in EUV channels (e.g., AIA at 211\,\AA) or in X-rays probing hotter plasma.
It will be interesting to investigate the role of these small-scale brightenings for the quiet-Sun X-ray emission in a future study.
For the sample campfire in \fig{f2}, we can conclude that it shows plasma at around 1\,MK and is thus also consistent with the observed campfires (item 6 in the introduction).

In investigating the temporal evolution of the plasma at the location of the transient brightening, we find that the enhancement in EUV radiation is caused by an increase in temperatures.
The plasma that is present at the location prior to this campfire gets heated and raises its temperature by more than a factor of ten from below the transition region to coronal temperatures.
This results in an expansion of the gas that goes along with a decrease in density.

Finally, we turn to the radiative flux from the campfire.
For this, we rebinned the intensity map to the plate scale of EUI (ca. 200\,km at 0.6\,AU) and found peak count rates around 1000\,DN per pixel and second.
When integrating this over the spatial extent and the lifetime of the brightening, the total number of counts was about $6{\times}10^5$\,DN.
To compare this to the values in \cite{campfires}, we had to convert the DN into detected photons (ca.\ 6.8\,DN per photon) and account for the exposure time (3\,s) being shorter than the image cadence (5\,s).
Then, for the sample campfire highlighted in \fig{f1}, we found a total of about 50\,000 photons.
This is in the range of the detected photons in observed campfires \cite[Fig.\,2 of][]{campfires}, as is expected for its comparatively large size and long lifetime in the upper range of the observed distribution.
Thus this sample model campfire is consistent with observations in terms of radiative output (item 7 in the introduction).

%%%%%%%%%%%%%%%%%%%%%%%%%%%%%%%%%%%%%%%%%%%%%%%%%%%%%%%%%%%%%%%%%%%%%%%%%%%%%
%
% In Berghmans et al 2021 in Fig.2 the "total photons" 
% are detected photons and they are simply added up 
% from the exposures. For a more proper calibration, 
% one would have to multiply this by 5s/3s 
% (5s cadence, 3s exposure).
%
% The (rough) conversion is 1000 photons in Fig.2  =  6800 DN
%
% Typically, peak counts in observed campfires are 500-1000 DN/pxl.
% This would be a few 100 DN/pxl/s.
%
% Simple count estimate from observations:}\\
%
% small ones
%
% counts:   500 DN/pxl            |
% size:       8 pxl               | total  = 12.000 DN 
% lifetime:   3 snapshots (15 s)  |        = 1700photons
%
% bigger ones
%
% counts:   1000 DN/pxl           |
% size:       32 pxl              | total  = 400.000 DN
% lifetime:   12 snapshots (60 s) |        =  50.000 photons
%
%%%%%%%%%%%%%%%%%%%%%%%%%%%%%%%%%%%%%%%%%%%%%%%%%%%%%%%%%%%%%%%%%%%%%%%%%%%%%

The fact that the sample transients we discuss in this model are on the larger side of the observed campfires is partly due to our selection criteria. We wanted to select a limited number of examples for a case-by-case study and concentrate on the larger events.
In addition, the spatial resolution in our numerical model is only a factor of four better than in the observations in terms of grid spacing or pixel size. The smallest, 1-pixel features identified by EUI as campfires might have magnetic sources in the photosphere that might not be resolved in our numerical model. Hence we might miss these smallest events in our model.

The seven campfires we have studied here have common observable signatures.
Just as the one example we discussed in detail above, they match the properties of the observed campfires summarized in items 1 to 7 in the introduction.
Their lifetimes can be seen from \fig{fs2}.
Their morphology, in particular length and aspect ratio, is illustrated in \figs{f3}, \ref{f4}, and \ref{fs1}, and their lifetimes are illustrated in \fig{fs2}.
From these figures it is also clear that they appear close to magnetic concentrations, but with some offset to the network patches.
 We illustrate the heights for only two cases (in \figs{f2} and \ref{f4}), but the others also occur at heights around 2\,Mm to 5\,Mm above the photosphere.
Here, we show data on temperature and radiative losses only for one case, but the others are of the same order of magnitude.
Based on this, we can conclude that the seven brightest or strongest coronal transients we find in our numerical model fit the properties of the observed campfires well.

\subsection{Magnetic field in and  around the coronal brightenings \label{S:part2}}

Having established that the transient coronal brightenings in our model match the properties of the observed campfires, the main question is what causes these brightenings in our model.
For this we investigate the magnetic field structure in and around these brighenings and, in particular, follow its temporal evolution.

For the analysis of the magnetic field for each of the seven events, we traced field lines crossing the brightening and followed this in time.
For each case, we randomly selected 15 to 20 seed points in the region of enhanced EUI 174 {\AA} emission when the campfire was close to its peak intensity (see Appendix\,\ref{S:appendix.lifetime}) and traced the magnetic field lines from these points in both directions. We chose the number of field lines in each case so that it was possible to follow them in the figures; we modified the locations and numbers of the seed points and found that this provides similar results. To capture the evolution of the magnetic field, we used the same seed points to trace field lines in the other snapshots to study the temporal evolution of the magnetic field topologies around the campfires.

\begin{figure*} 
\centering {\includegraphics[width=\textwidth]{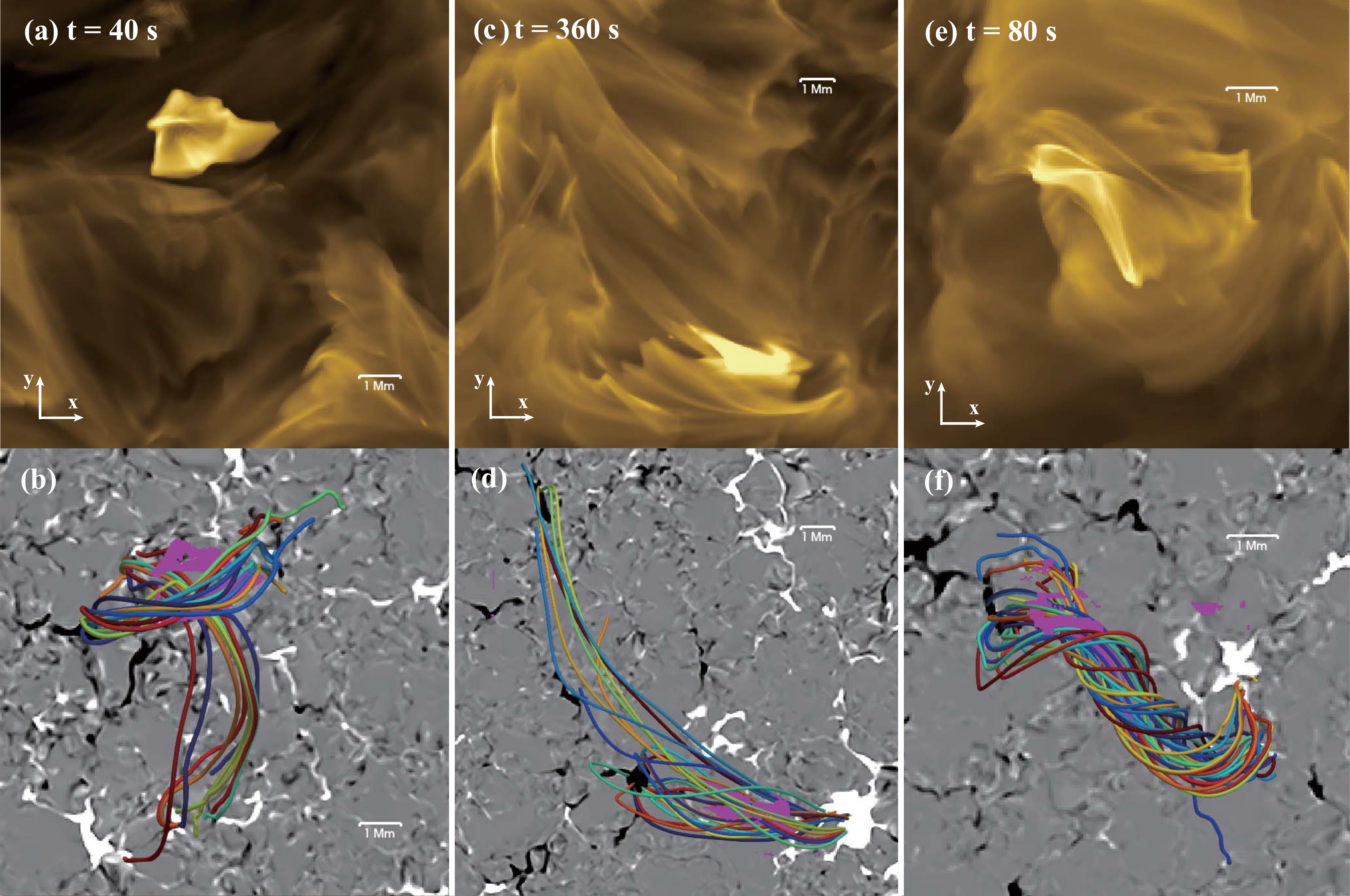}} 
\caption{Campfires and their relation to the magnetic field.
Panel (a) displays the synthetic EUI 174 {\AA} image of a small field-of-view around a transient brightening close to its maximum intensity (the time given in the panel label; see also light curves in \fig{fs2}).
The x and y directions are marked and a bar indicates a scale of 1\,Mm.
The intensity scale is logarithmic.
Panel (b) shows the vertical magnetic field in the photosphere in the same field of view and at the same time as in panel (a).
The magnetogram is saturated at $\pm$350\,G.
In purple we show a mask of the bright EUV emission in panel (a).
The differently colored lines show field lines that  pass through the volume of the transient brightening (cf. \sect{S:part2}).
Panels (c/d) and panels (e/f) show the same thing as panels (a/b), just for different campfires.
These three examples represent the groups (i) to (iii) defined in \sect{S:part2}.
More examples are shown in the same format in \figs{f4} and \ref{fs1}.
Animations for these three cases are available online.
See \sect{S:part2}.
%
%{\newline\color{red} Add thin white spaces again between (a/b), (c/d) and (e/f) to have a better distinction between the cases.}
%
} \label{f3}
\end{figure*}

We found that the magnetic field structures of the seven identified campfires can be categorized into three groups. For each group we display an example in \fig{f3} showing one snapshot in the 174\,\AA\ band while the campfire is almost brightest side by side with the selected magnetic field lines. 
%The three groups are as follows:\todo{ Please consider reformatting.}
%
%\begin{description}
%    \item[(i)]   crossing bundles of field lines (three cases; sample in \fig{f3}a/b),
%    \item[(ii)]  forking bundles of field lines (three cases; \fig{f3}c/d), and
%    \item[(iii)] highly twisted flux ropes (one case; \fig{f3}e/f).
%\end{description}
{The three groups are (i) crossing bundles of field lines (three cases; sample in \fig{f3}a/b), (ii) forking bundles of field lines (three cases; \fig{f3}c/d), and (iii) highly twisted flux ropes (one case; \fig{f3}e/f).}

\begin{figure*} 
\centering {\includegraphics[width=18cm]{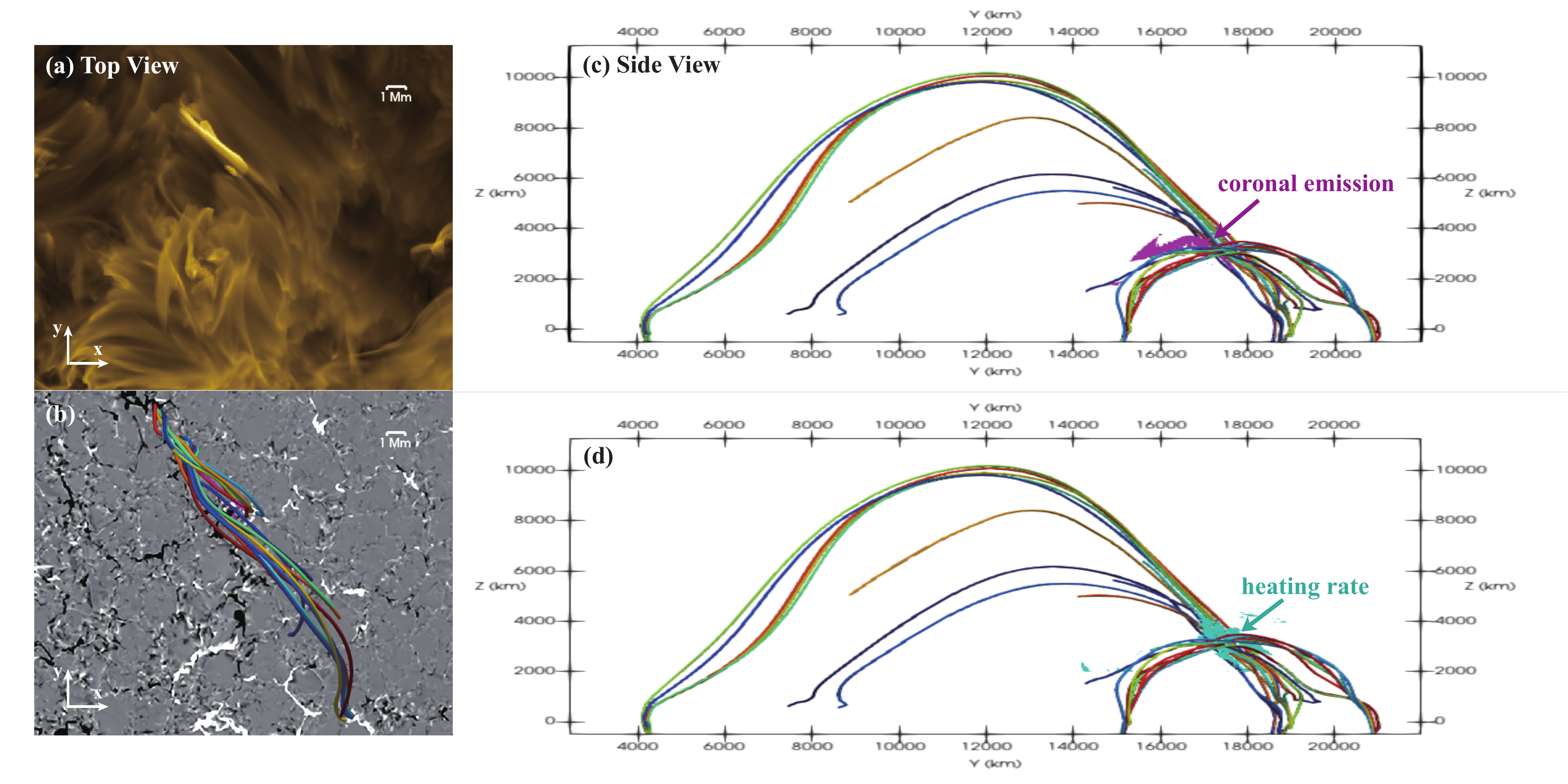}} 
\caption{Spatial appearance of campfires.
Panels (a/b) show another example of a campfire in the same format as in \fig{f3} (at time $t{=}260$\,s).
In panel (c) and (d), we show a projection of the field lines in panel (b) onto the vertical $y$-$z$ plane.
All the field lines are mostly in a vertical plane at a ca. 30$^\circ$ angle to the y-direction.
Therefore we scaled the side view in panels (c/d) so that the field lines appear at the approximately correct aspect ratio of the horizontal to vertical extent.
In panel (c) we overplot a mask of the EUI 174\,\AA\ emission in purple, and in panel (d) the heating rate is overplotted (resistive plus viscous) in cyan.
An animation of panel (c) is available online.
See \sect{S:part2}.
} 
\label{f4}
\end{figure*}

The brightenings in group (i) are associated with two bundles of field lines.
In the initial stages, these bundles are almost perpendicular to each other. The enhanced EUI 174 {\AA} emission predominantly originates from the interface between the two bundles (\fig{f3}a/b).
In this case, the field-line bundle running along the y-direction is rising.
Following the interaction between the two bundles, only the field lines of the bundle in the x-direction remain (see animation associated with \fig{f3}a/b).
This is indicative of reconnection, and in response to the heating the plasma gets hot enough to radiate in the 174\,\AA\ band.
The other two examples of this case are shown in \fig{f4}a/b, \fig{fs1}e/f, and the associated animations.

The brightenings of group (ii) are associated with bundles of forking field lines.
Before their interaction, the two bundles originate from one common magnetic patch in the photosphere. The other ends of the bundles are rooted at two separate magnetic patches, leading to an overall appearance of a forking structure (see \fig{f3}c/d).
During the evolution, again illustrated in animation with the figure, the heating and subsequent brightening appear at the location where the two field-line bundles are separating, that is, form the fork.
As in group (i), after the brightening, only one of the bundles of field lines remains.
The other two examples of this group are shown in \fig{fs1}a/b and c/d.

The events in group (i) and (ii) are quite similar in terms of magnetic topology.
Firstly, after each of these six events, we found that one bundle of magnetic field lines is almost completely eroded once the campfire disappears (see animations with the figures).
This also underlines that reconnection is the energy source for the brightening.
Secondly, the reconnection between the field lines involved is occurring at small angles of 90\,degrees or less, that is, the reconnecting field lines {are nearly perpendicular to each other or} point roughly in the same direction.
%
%
%The second example, shown in \fig{f3}(c)-(d), belongs to the second group. This campfire is the one presented in \fig{f1} and \fig{f2}. It reveals a flame-like morphology, and there are also two bundles of field lines around it. These field lines originate from a common magnetic element in the photosphere. The other footpoints of these field lines appear to be located at two different patches of magnetic features, thus forming mainly two bundles of fied lines or two arcades. The far-reaching arcade is also higher. The enhanced EUI 174 {\AA} emission appears to be located at the top of the lower arcade, which is also the interface between the two arcades. We present another four cases that are similar to the two cases mentioned above in \fig{f4} and \fig{fs1}. For each of these six cases, we found that one bundle of magnetic field lines is almost completely eroded when the campfire disappears (see the online animations S2 and S3).
%
%
This suggests that the campfires are most likely caused by component magnetic reconnection between different magnetic structures \citep[e.g., ][]{Sonnerup1974}, a process otherwise mostly considered in open magnetic configurations such as coronal holes \cite[e.g.,][]{Tenerani2016,Viall2020}.
This is unlike anti-parallel reconnection that was frequently seen in {recent} MHD simulations of 
%transient brightenings
{explosive events and UV bursts} \citep[e.g.,][]{Innes2015, Ni2016, Priest2018,Syntelis2019,Peter2019, Ni2020,Danilovic2017,Hansteen2017, Hansteen2019}, where the angle between the directions of reconnecting magnetic field lines is generally about 180 degrees.

%>>>>>>>>>>>>>>>>>>>>>>>>>>>>>>>>>>>>>>>>>>>>>>>>>>>>>>>>>>>>>>>>>>>>>>>>>>>>>>
%<<<<<<<<<<<<<<<<<<<<<<<<<<<<<<<<<<<<<<<<<<<<<<<<<<<<<<<<<<<<<<<<<<<<<<<<<<<<<<

For one example from groups (i) and (ii), we take a closer look at the relation of reconnection, heating, and brightening in the EUV.
The campfire shown in \fig{f4}a exhibits an elongated loop-like morphology. The magnetic field lines around the campfire show a very high degree of overlapping when viewed from the top (\fig{f4}b). To better illustrate the magnetic field structures around the campfire, we show a projection onto the vertical $yz$--plane.
In this projection, the field lines from the two bundles show a crossing under small angles, as is common for the other group (i) and (ii) events.
At the location where the two field-line bundles cross, we find an enhanced heating rate (resistive plus viscous), which underlines the presence of reconnection heating the plasma.
This is illustrated in \fig{f4}d and underlines the presence of component reconnection heating the plasma.
The emission in the 174\,\AA\ band will not be identical to the location of the heat input, of course. For example, if one of the two bundles of field lines has a lower density than the other, then we will mostly see the enhanced emission on the bundle where the (pre-event) density was higher.
In the example in \fig{f4}, this is the case for the lower-lying bundle.
Because the field lines are, very roughly, semicircular (see \fig{f4}) and the typical distance between opposite magnetic polarities in the quiet-Sun network is on the order of a few granules, we can typically expect these closed field lines to reach not much higher than 5\,Mm \cite[e.g.,][]{Jendersie2006}.
Consequently, the campfires also mostly do not exceed this height, and the seven examples we analyze here are occurring at heights between 2\,Mm and 5\,Mm.
This height range for the campfires roughly matches the height range where the normal quiescent coronal emission in the 174\,\AA\ band originates.

From this we conclude that for the events from groups (i) and (ii), component reconnection leads to heating of plasma in the vicinity of the reconnection region of the interacting field-line bundles.
This causes the region near the top of one of the field-line bundles to brighten.
This supports one of the interpretations given in the observations by \cite{campfires} to explain why the length of the campfires is smaller than twice their height above the photosphere.

%the magnetic field lines, EUI 174 {\AA} emission, and heating rate (the sum of resistive and viscous heating) onto the $yz$--plane. We can see that the coronal brightening is located at a height of 3--4 Mm, indicating that the campfire is a low-corona phenomenon. There are two separate field-line bundles (or arcades) around the coronal brightening, and the heating rate at the interface is enhanced. We also found that both arcades are roughly semi-circular, and the emission is constrained at the top of the lower-lying arcade. We found that this magnetic field structure changes to a single arcade when the campfire disappears (see the online animation S5). This suggests that interaction between the two arcades, i.e., component magnetic reconnection, heats the plasma and generates the coronal brightening. Similar to the case shown in \fig{f2}, the plasma at the reconnection site is heated to $\sim$3 MK, and the EUI 174 {\AA} emission is depleted after reconnection. This also explains the slight offset between the regions of enhanced heating and enhanced EUI 174 {\AA} emission.

The single event in group (iii) is quite different in nature and obviously associated with an untwisting magnetic flux rope.
At the location of the brightening both before and during the event, the field lines visualize a flux rope (\fig{f3}e/f). 
The temporal evolution of the magnetic field structure shows the relaxation of this highly twisted flux rope
%by initially at least 1 to 2 windings from end to end 
(see animation associated with the figure).
At the end of the brightening in the 174\,\AA\ channel, the bundle of field lines is largely, but not completely, untwisted.
It is possible that component magnetic reconnection between internal adjacent field lines within the flux rope \citep[e.g.,][]{Xing2020} heats the plasma to 1 MK or more and also results in a reduction of twist.
Due to the small number of events investigated in this study, we cannot determine whether this mechanism makes a substantial fraction of campfires.
%
%Based on the small number of events investigated in this study, we would be inclined to conclude that this untwisting of flux ropes might be responsible for only a small but non-negligible fraction of campfires.

%>>>>>>>>>>>>>>>>>>>>>>>>>>>>>>>>>>>>>>>>>>>>>>>>>>>>>>>>>>>>>>>>>>>>>>>>>>>>>>
%<<<<<<<<<<<<<<<<<<<<<<<<<<<<<<<<<<<<<<<<<<<<<<<<<<<<<<<<<<<<<<<<<<<<<<<<<<<<<<

%==============================================================================
\section{Discussion\label{S:discussion}}
%==============================================================================

Our results derived from the (synthesized) coronal emission seen in the transient brightenings show a good match to the campfires as reported in the observations by \cite{campfires}.
In \sect{S:part1}, we detail how the seven items listed in the introduction are recovered by our model.
This suggests that the mechanism(s) driving the transients in our model might resemble the physics governing the observed campfires, even though one should remember that the Sun might still choose a different path.

\begin{figure}
\centering
\includegraphics[width=8.8cm]{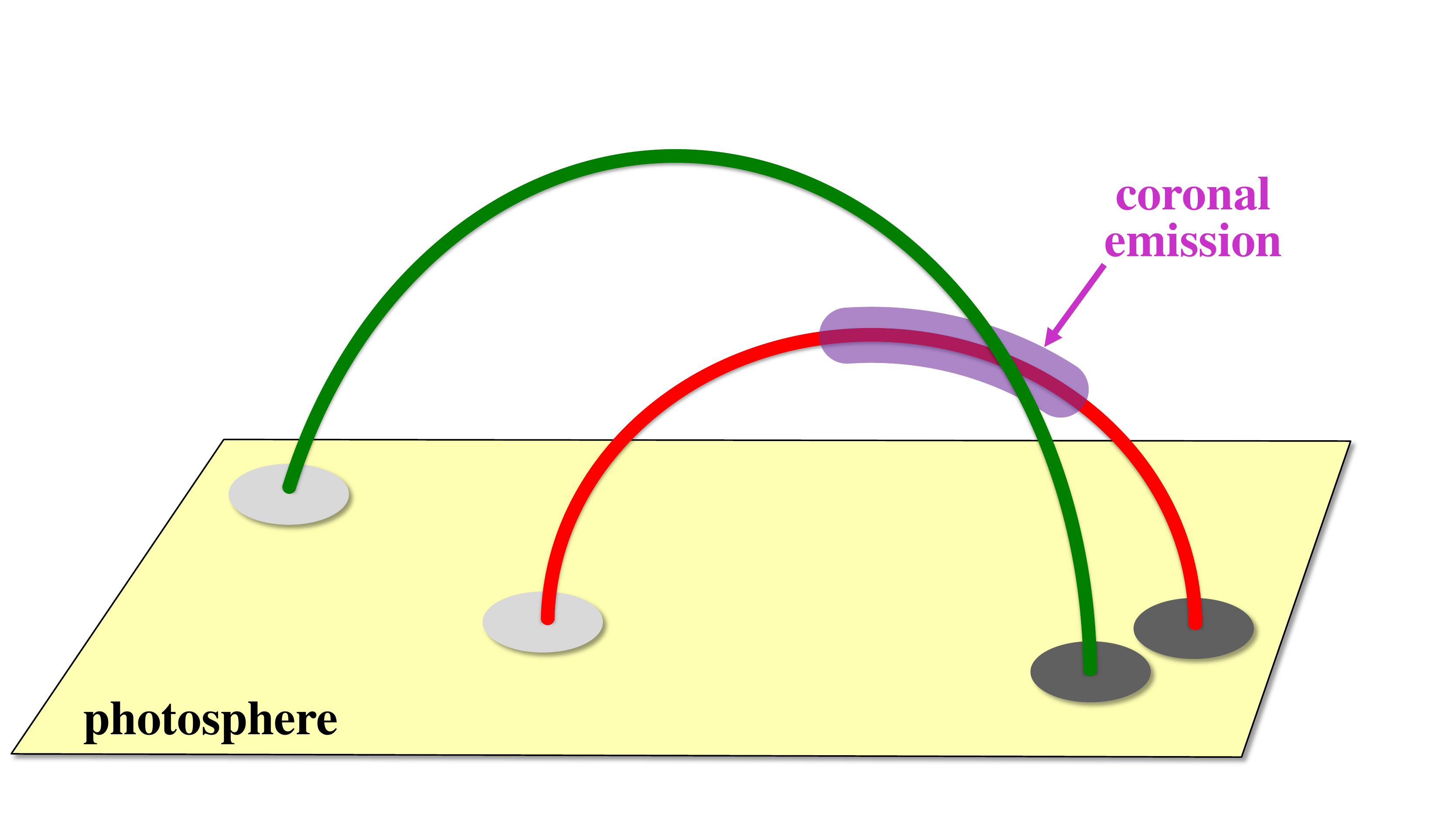}
\caption{Magnetic structure of a campfire.
This illustrates the majority of the campfires in our model, i.e., those from groups (i) and (ii) defined in \sect{S:part1}.
The light and dark gray areas represent patches of opposite magnetic polarity in the photosphere where the two interacting bundles of field lines (red and green) are rooted.
Following the reconnection, coronal emission is seen along a loop-like stretch (purple).
Essentially, this follows the structure found in the model for the case shown in \fig{f4}c.
See \sect{S:discussion}.
}\label{f6}
\end{figure}

In our model, we find that in the majority of cases, the energization mechanism driving the transient brightening is the component reconnection in the coronal part of interacting bundles of field lines, viz. coronal loops.
This is illustrated in \fig{f6}.
While heating is mostly restricted to the region where the field-line bundles interact, the brightening is visible through the heated plasma on the interacting loop bundles.
In the cartoon in \fig{f6}, we imply that the density, before interaction, was higher on one of the two field-line bundles, and consequently this dominates the emission pattern and forms a loop-like structure stretching away from the reconnection region. {At the lower parts of the field-line bundles, the densities are high so that the released energy is not sufficient to heat the plasma there to $\sim$1 MK. As a result, the EUI 174\,\AA\ emission is confined at the higher parts and it does not reach the footpoints of the field-line bundles.}
This magnetic and emission structure is quite similar to the transient brightening in the MHD model shown in \fig{f4}c.

%>>>>>>>>>>>>>>>>>>>>>>>>>>>>>>>>>>>>>>>>>>>>>>>>>>>>>>>>>>>>>>>>>>>>>>>>>>>>>>
%<<<<<<<<<<<<<<<<<<<<<<<<<<<<<<<<<<<<<<<<<<<<<<<<<<<<<<<<<<<<<<<<<<<<<<<<<<<<<<

This type of magnetic interaction has been found before for active regions and flaring loops.
The rearrangement of tangled loops has been found to be responsible for the heating of some compact hot active-region loops \citep{Reale2019} and post-flare loops \citep{Parenti2010}.
Also examples of cool loops that stay well below 0.1\,MK have been found to be heated in response to interacting flux bundles \citep{Li2014}.
The processes in these loops would correspond to the campfires of group (i) and (ii) as defined in \sect{S:part1}. 
The event in group (iii) might be more comparable to flux ropes as found in solar flares and coronal mass ejections characterized by highly twisted field lines \citep[e.g.,][]{Shibata2011,Cheng2017}.
In this study, we find that there is one common aspect of these active-region features and the campfire events: The energy to heat these structures is deposited in the upper (coronal) part of the features, in some cases near the loop apex.
In the case of the campfires, this energizing near the apex is consistent with the observation of them occurring at heights significantly above the photosphere.
Furthermore, it is worth mentioning that component reconnection can generate mini-jets in prominence structures \citep{Chen2017,Chen2020,Antolin2021}.

%Our analysis results suggest that campfires observed in the quiet-Sun corona are likely caused by component magnetic reconnection between interacting magnetic loop structures or internal field lines within flux ropes. \citet{Li2014} reported the conversion from mutual helicity to self-helicity as a result of reconnection between two crossing loops. This process, though reported in an active region, might also be the case for the quiet-Sun campfires generated by interacting loop structures in our model. In addition, the rearrangement of tangled loops has been found to be responsible for heating of some compact hot active-region loops \citep{Reale2019} and post-flare loops \citep{Parenti2010}. Observations of solar flares and coronal mass ejections have frequently revealed the existence of magnetic flux ropes characterized by highly twisted field lines \citep[e.g.,][]{Cheng2017}. Our results suggest that similar magnetic field topologies could also be found in the quiet Sun, and that similar physical processes are likely responsible for many transient small-scale brightenings in the quiet-Sun corona. 

In contrast to these campfires, for many if not most small-scale transients on the Sun, the energy is thought to be supplied through reconnection at the footpoints that accompanies flux emergence and cancellation.
This applies to Ellerman bombs \citep[e.g.,][]{Georgoulis2002,Schmieder2014}, UV bursts \citep[e.g.,][]{Peter2014,Tian2016,Tian2018,Li2018}, and spicules \citep[e.g.,][]{Yurchyshyn2013,Samanta2019}.
Small-scale coronal brightenings in the core of active regions are closely related to flux emergence \citep{Tiwari2019}.
Even coronal loops, both regular and flaring, might be heated through reconnection at or near their loop footpoints \citep{Chitta2017sunrise,Chitta2018}.
A common feature of these events is flux emergence and/or cancellation.
Unfortunately, for the campfire observations by \cite{campfires}, no high-resolution magnetic field observations are available to establish if campfires are related to flux emergence or cancellation. %
In our model, we do not see signatures of flux emergence or cancellation {in the photosphere around the campfires}.
Instead, the loops, or field-line bundles, are driven by horizontal motions in the photosphere and they then interact at coronal heights through component reconnection.
Future magnetic field observations are required to show if this prediction of our model is correct, namely that campfires are not clearly related to flux emergence and cancellation.
This also underlines the need for measurements of the coronal magnetic field \citep[e.g.,][]{LiW2015,Yang2020a, Yang2020b, Landi2020}.

%The coronal brightenings are often located around the top of the lower-lying loops. Because of this, the coronal brightenings often reveal an apparent offset from the network magnetic elements when viewed from the top. It is worth mentioning that we found no obvious signature of flux emergence or cancellation in the photospheric magnetograms during the occurrence of these coronal campfires. This is very different from many other types of small-scale transient events such as Ellerman bombs, UV bursts and spicules, which are formed below the corona and flux cancellation is often observed \citep[e.g.,][]{Georgoulis2002,Peter2014,Tian2018,Samanta2019}. As the campfires are caused by component magnetic reconnection at coronal heights, they do not cause evident changes in the photospheric magnetograms. Hence, direct measurements of the temporal evolution of vector magnetic field above the photosphere are necessary to understand the generation mechanisms of campfires.

Component magnetic reconnection naturally occurs in a scenario of braiding magnetic field structures \citep{Parker1983, Parker1988}. 
{Pioneering 3D MHD numerical experiments about magnetic braiding were already performed in the 1990s \citep[e.g.,][]{Galsgaard1996}. Since then, significant progress has been made in realistic simulations of coronal heating caused by footpoint braiding \citep[e.g.,][]{Gudiksen2002,Bingert2011,Hansteen2015,Rempel2017}. Properties of small-scale heating events caused by the current dissipation in the corona have also been explored through 3D MHD simulations \citep[e.g.,][]{Bingert2013, Guerreiro2015, Guerreiro2017}.}
However, direct possible observational evidence of braiding is sparse  \citep{Cirtain2013,Huang2018}. 
According to our model, in many of these campfires, the angles between reconnecting field lines are small, with field lines generally pointing in a similar direction. {Thus we might consider campfires as indirect evidence for this process.}

In principle, an increase in the EUV emission in a particular wavelength band or spectral line originating at a given temperature can occur for several reasons: The plasma is heated from lower temperatures or is cooling from higher temperatures, or the density increases, for example, due to a flow into the structure or by compression through the magnetic field \cite[e.g.,][]{blinkers}.
The cause of EUV brightenings in the transition region has been debated, with
a slight preference for density increase \cite[e.g.,][]{Bewsher2003}.
For the coronal campfires, we see in our model that we can safely conclude that they are solely due to the heating of plasma on the reconnecting bundles of field lines.

Finally, we turn to the energetics of the campfire events in our model.
{We estimated the total dissipated energy (viscous plus resistive heating) during the campfire shown in \fig{f4}. We first selected the snapshot when the campfire reached its peak intensity. Then, we integrated the heating rate at all pixels within the cyan patch in \fig{f4}, where the heating rate is significantly enhanced. The total energy was found to be on the order of $10^{27}$ ergs. We also examined the other events and found similar results. }Thus, the campfires might be categorized as microflares in terms of energy. 
{The energy is nearly two orders of magnitude smaller than that of a typical UV burst \citep{Peter2014}, which likely occurs at a lower height of $\sim$1 Mm above the photosphere \citep{Tian2018,Chen2019}.}
We identified seven events over the span of 16 minutes within an area of 50$\times$50 Mm$^{2}$, which corresponds to an average energy flux of roughly 3$\times$10$^{5}$\,erg\,cm$^{-2}$\,s$^{-1}$. This supply is comparable to the canonical amount of energy required to heat the quiet-Sun corona \citep{Withbroe1977}. Furthermore, additional campfire-like events would be identified if we were to lower the threshold for event detection. Considering this, we conjecture that campfires should play an important role in the energization of the quiet-Sun coronal plasma. Some observational studies suggest that bursty events in the quiet-Sun corona do not play a dominant role in the heating of the corona \cite[e.g.,][]{Aschwanden2000,Chitta2021}. 
{Whether campfire-like events can maintain a hot corona on the real Sun is still unclear.}
Further observational studies, particularly with the superior spatial resolution of EUI/HRT on Solar Orbiter, and a more thorough investigation through numerical experiments are needed to settle this issue.

\section{Conclusions\label{S:conclusions}}
%==============================================================================

To understand the processes underlying the campfires recently observed with EUI \citep{campfires}, we employed a numerical 3D MHD model of the quiet Sun using the MURaM code.
The model self-consistently creates the typical magnetic patterns of supergranulation with a million K hot corona above.

The transients we find in 174\,\AA\ emission synthesized from our numerical model show properties that match those of observed campfires.
They match in terms of lifetime, size, aspect ratio, temperature, and radiative losses; furthermore, they occur at the same height adjacent to, but not directly above, bright patches of the chromospheric network.
With this, the transients in our model are in line with the seven items listed in the introduction that summarize the observational findings of campfires by \cite{campfires}.

%To understand the generation mechanisms of the transient coronal brightenings or campfires observed by the EUI instrument on board Solar Orbiter, we have performed realistic radiation MHD simulations using the MURaM code, and obtained a self-consistent and time-dependent model of the quiet Sun. A stable corona at an averaged temperature of 10$^{6}$ K is maintained in our model. We synthesized the coronal emission in the EUI 174 {\AA} passband and identified seven transient coronal brightenings that are similar to the campfires observed during the first perihelion of Solar Orbiter.

The MHD model contains the full information of the self-consistent 3D evolution of the magnetic field.
Unlike in the observations, we can thus investigate the details of the magnetic processes in and around the campfires.
From this we conclude that the campfires are mostly caused by component reconnection between interacting bundles of magnetic field lines.
The interaction of the bundles and thus the heating to 1\,MK or more and the subsequent brightening in the EUV typically occurs at heights of up to 5\,Mm and is slightly offset (horizontally) to the magnetic concentrations that define the bright patches of the chromospheric network.
In one out of seven cases, we found that untwisting of a flux rope is responsible for the heating of loop-like features, which in EUV very much resembles the other cases due to component reconnection.
We do not find obvious signatures of flux emergence or cancellation in the photosphere in any cases.
Still, in our model we find that these transient brightenings might supply a sufficient amount of energy to heat the quiet-Sun corona.

Future investigations of the statistical properties of these transient events will have to show if what we found in the model is indeed the same type of feature as the observed campfires.
The good match between observed and modeled properties, for a limited number of cases, makes us rather confident that this will be the case.

%==============================================================================
%==============================================================================
%==============================================================================
%==============================================================================
\begin{acknowledgements}
This work is supported by NSFC grants 11825301, 11790304 and 12073004, the Strategic Priority Research Program of CAS (grant no. XDA17040507), and the Max Planck Partner Group program. Y.C. also acknowledges partial support from the China Scholarship Council and the International Max Planck Research School (IMPRS) for Solar System Science at the University of G\"ottingen during his stay at MPS. Solar Orbiter is a space mission of international collaboration between ESA and NASA with contributions from national agencies of ESA member states. EUI was conceived by a multinational consortium and proposed in 2008 under the scientific lead of Royal Observatory of Belgium (ROB) and the engineering lead of Centre Spatial de Li\`ege (CSL). We thank Dr. L. P. Chitta and Feng Chen for helpful discussion. This project has received funding from the European Research Council (ERC) under the European Union’s Horizon 2020 research and innovation programme (grant agreement No. 695075). We gratefully acknowledge the computational resources provided by the Cobra supercomputer system of the Max Planck Computing and Data Facility (MPCDF) in Garching, Germany. D.P. would like to thank A. Irwin (Free-EoS) and V. Witzke (MPS-ATLAS) for the fantastic open-source packages they provide.
\end{acknowledgements}

\bibliography{refs}
\bibliographystyle{aa}

\begin{appendix}

\section{Additional examples of campfires}

Our study investigated the seven strongest (viz. brightest) campfire events  we found in our model.
Based on their magnetic structure, we classified these into three groups (see \sect{S:part2} for the definition).
In the main text, we show one example for each case in \fig{f3} and a further example in \fig{f4}.
Here we show the other three examples in the same format as in \fig{f3}.

%>>>>>>>>>>>>>>>>>>>>>>>>>>>>>>>>>>>>>>>>>>>>>>>>>>>>>>>>>>>>>>>>>>>>>>>>>>>>>>
\begin{figure*} 
\centering {\includegraphics[width=\textwidth]{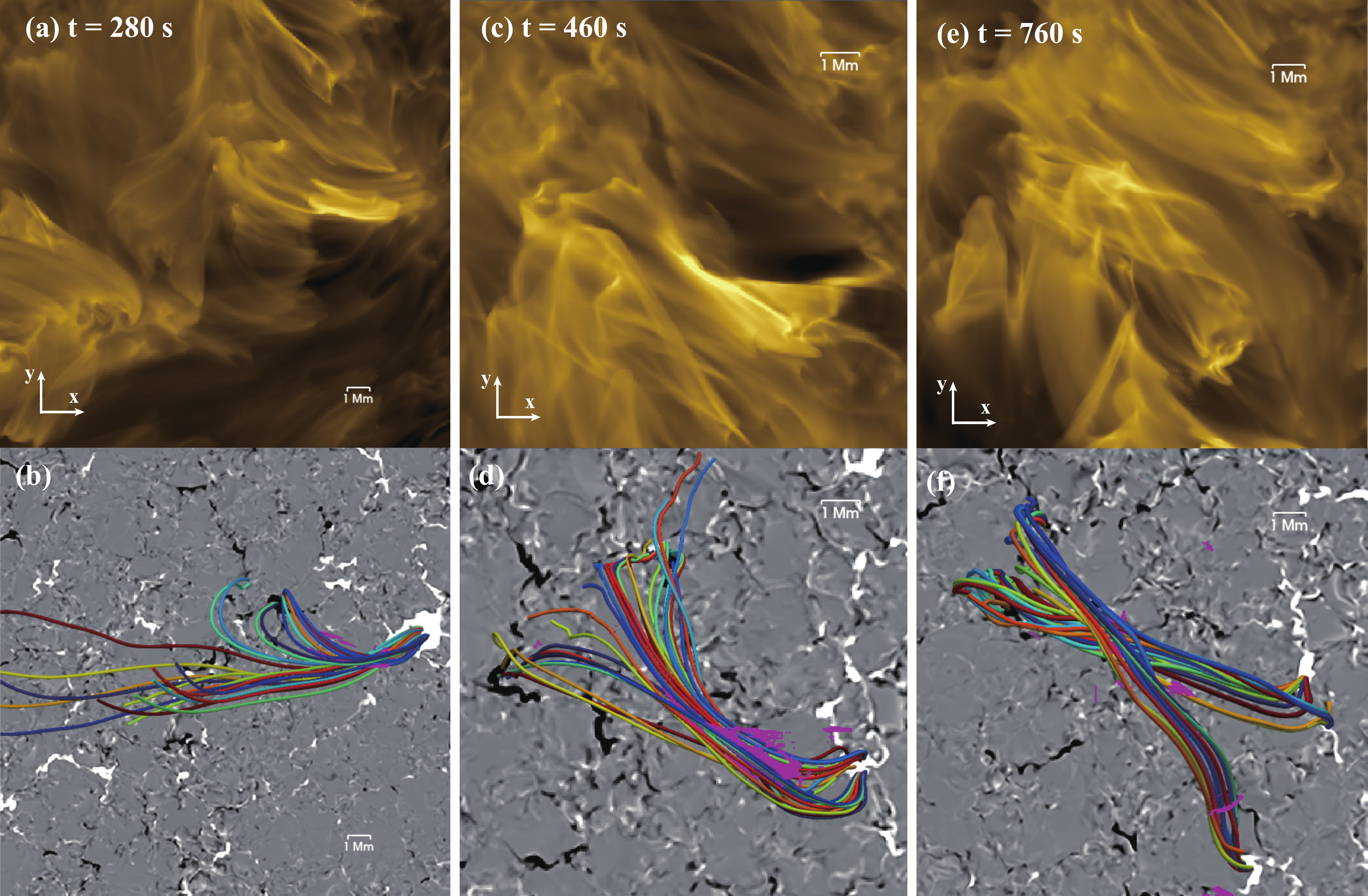}} 
\caption{Relation of campfires to the magnetic field for three more examples. Same as \fig{f3}, but for the cases not shown in the main paper in \figs{f3} and \ref{f4}. Animations for each example are available online.
%
%{\newline\color{red} Add thin white spaces again between (a/b), (c/d) and (e/f) to have a better distinction between the cases?}
%
} \label{fs1}
\end{figure*}
%<<<<<<<<<<<<<<<<<<<<<<<<<<<<<<<<<<<<<<<<<<<<<<<<<<<<<<<<<<<<<<<<<<<<<<<<<<<<<<

\section{Lifetimes of the campfires}\label{S:appendix.lifetime}

\begin{figure}
\centering
\includegraphics[width=7cm]{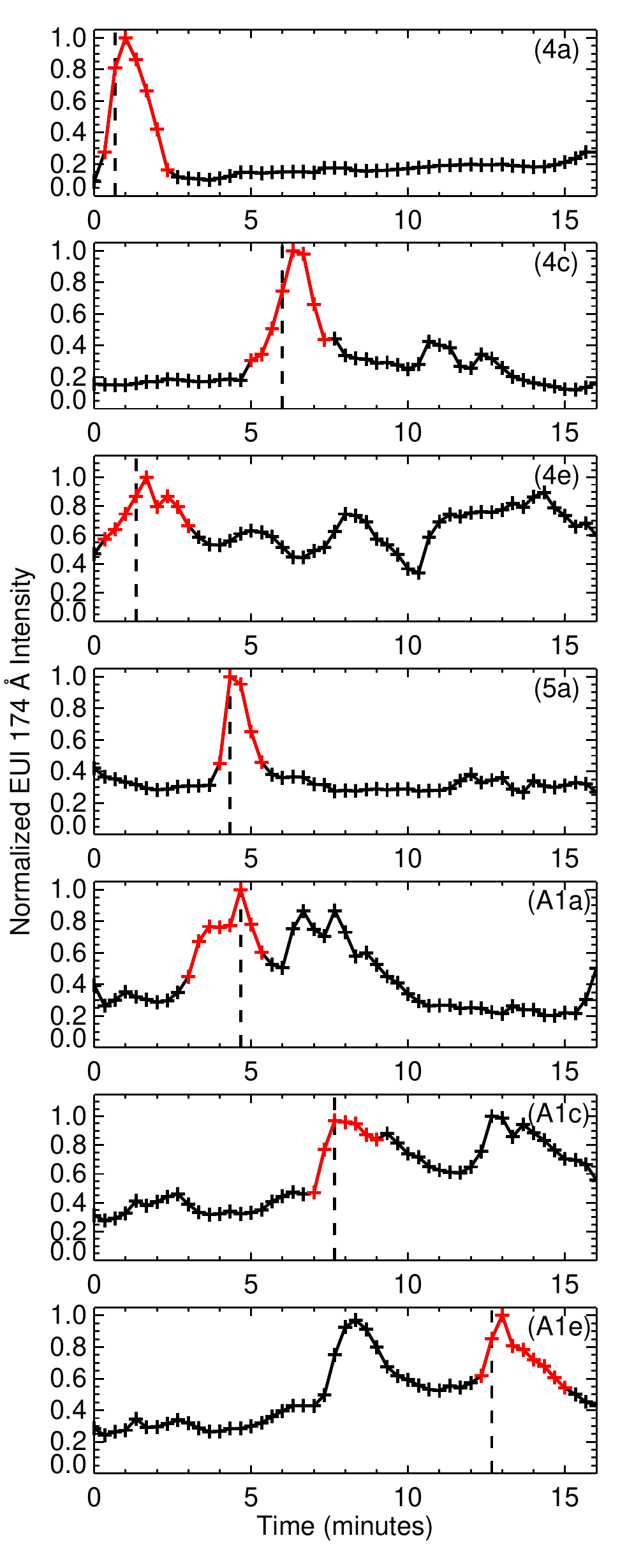}
\caption{Light curves of modeled campfires.
The panels show the temporal evolution in the EUI 174\,\AA\ band as synthesized from our model.
These campfires and their relation to the magnetic field are shown in detail in \figs{f3}, \ref{f4}, and \ref{fs1}, as referred to by the labels in the top right of each panel.
The red parts correspond to the periods when the campfires appear in the EUI 174 {\AA} images and the vertical dashed lines indicate the times of the snapshots in \figs{f3}, \ref{f4}, and \ref{fs1}.
See Appendix \ref{S:appendix.lifetime}.
}\label{fs2}
\end{figure}

For each of the seven campfires in our model, we examined the temporal variation of the EUI 174\,{\AA} emission.
For this, we added up the intensity at full resolution over a small rectangle that contains the campfire.
The resulting light curves for the seven cases are displayed in \fig{fs2}.
These light curves are not sensitive to the size of the rectangle because the campfires we selected dominate the 174\,\AA\ emission locally.
We also marked the times of the snapshots displayed in \figs{f3}, \ref{f4}, and \ref{fs1} that are close to peak intensity.

We used these light curves to estimate the lifetimes of the campfires.
For this, we considered those snapshots (viz. times) as being part of the campfire for which the coronal brightening meets the criteria to identify campfires as described in \sect{S:part1}.
Some peaks in the curves were not identified as campfires because the spatial extent of the respective brightenings is smaller than our size criterion for campfire identification (cf.\ \sect{S:part1}). 
Based on this, we indicated the duration of the campfires (in red).
Typically, the lifetimes are below 3 minutes.

%>>>>>>>>>>>>>>>>>>>>>>>>>>>>>>>>>>>>>>>>>>>>>>>>>>>>>>>>>>>>>>>>>>>>>>>>>>>>>>

\end{appendix}

\end{document}